\documentclass[letterpaper, 10 pt, conference]{ieeeconf}  % Comment this line out if you need a4paper
\IEEEoverridecommandlockouts                              % This command is only needed if 
\usepackage{tabularx,xtab,array,float}
\usepackage{epsfig,epstopdf,wrapfig,psfrag}
\usepackage{amsmath,amsfonts,amssymb,graphicx,graphics}
\usepackage{tikz}
\usepackage{balance,physics,stackengine}
\usepackage[normalem]{ulem}
\usepackage{hyperref}

\makeatletter
\newcommand*{\encircled}[1]{\relax\ifmmode\mathpalette\@encircled@math{#1}\else\@encircled{#1}\fi}
\newcommand*{\@encircled@math}[2]{\@encircled{$\m@th#1#2$}}
\newcommand*{\@encircled}[1]{%
\tikz[baseline,anchor=base]{\node[draw,circle,outer sep=0pt,inner sep=.2ex] {#1};}}
\makeatother

\newcommand\copyrighttext{%
  \footnotesize \textcopyright \the\year{} IEEE.  Personal use of this material is permitted.  Permission from IEEE must be obtained for all other uses, in any current or future media, including reprinting/republishing this material for advertising or promotional purposes, creating new collective works, for resale or redistribution to servers or lists, or reuse of any copyrighted component of this work in other works.}

\newcommand\copyrightnotice{%
\begin{tikzpicture}[remember picture,overlay]
\node[anchor=south,yshift=10pt] at (current page.south) {\fbox{\parbox{\dimexpr\textwidth-\fboxsep-\fboxrule\relax}{\copyrighttext}}};
\end{tikzpicture}%
}

\newtheorem{remark}{Remark}

\begin{document}
\title{\LARGE \bf Propeller Motion of a Devil-Stick using Normal Forcing}
\author{Aakash~Khandelwal, and~Ranjan~Mukherjee,~\IEEEmembership{Senior Member}% <-this % stops a space
\thanks{This work was supported by the National Science Foundation, under Grant CMMI-2043464}%
\thanks{The authors are with the Department of Mechanical Engineering, Michigan State University, East Lansing, MI 48824, USA
{\tt\footnotesize khande10@egr.msu.edu}, {\tt\footnotesize mukherji@egr.msu.edu}}% <-this % stops a space
}
	
\maketitle
\thispagestyle{empty}
\pagestyle{empty}
\copyrightnotice

\begin{abstract}
The problem of realizing rotary propeller motion of a devil-stick in the vertical plane using forces purely normal to the stick is considered. This problem represents a nonprehensile manipulation task of an underactuated system. In contrast with previous approaches, the devil-stick is manipulated by controlling the normal force and its point of application. Virtual holonomic constraints are used to design the trajectory of the center-of-mass of the devil-stick in terms of its orientation angle, and conditions for stable propeller motion are derived. Intermittent large-amplitude forces are used to asymptotically stabilize a desired propeller motion. Simulations demonstrate the efficacy of the approach in realizing stable propeller motion without loss of contact between the actuator and devil-stick.
\end{abstract}


\section*{Nomenclature}
\begin{tabularx}{\columnwidth}{lX}
$g$ & acceleration due to gravity, (m/s$^2$)\\
$h_x, h_y$ & Cartesian coordinates of the center-of-mass of the devil-stick, (m) \\
$\ell$ & length of the devil-stick, (m)\\
$m$ & mass of the devil-stick, (kg)\\
%$q$ & generalized coordinates, $q \triangleq \begin{bmatrix} h_x & h_y & \theta \end{bmatrix}^T$ \\
$r$ & distance of point of application of force from the center-of-mass of the devil-stick, (m) \\
$F$ & normal force applied on the devil-stick, (N)  \\
$I$ & impulse of impulsive force applied on the devil-stick, (Ns) \\
$J$ & mass moment of inertia of the devil-stick about its center-of-mass, (kgm$^2$) \\
$\theta$ & orientation of the devil-stick, measured positive counterclockwise with respect to the horizontal axis, (rad) \\
% $(.)^-$, $(.)^+$ & variable $(.)$ immediately before and after an event where there is a discontinuous jump in its value \\
\end{tabularx}

\section{Introduction} \label{sec:introduction}

We revisit the problem of realizing rotary \emph{propeller} motion of a devil-stick using continuous-time forcing. This problem is one of nonprehensile manipulation since the devil-stick is manipulated without grasping, and is subject to a unilateral constraint at its point of contact with the manipulator. 
Among others, nonprehensile manipulation has the advantage of being able to control more degrees-of-freedom (DOFs) than that of the manipulator \cite{lynch_dynamic_1999, ruggiero_nonprehensile_2018}. 
Nonprehensile manipulation tasks may involve either continuous or intermittent contact \cite{ruggiero_nonprehensile_2018, khandelwal_nonprehensile_2023} between the object and manipulator. The work in this paper belongs to the former class of problems.

Propeller motion of a devil-stick has been previously studied in \cite{kawaida_feedback_2003, nakaura_enduring_2004, nakamura_enduring_2009, shiriaev_generating_2006, aoyama_realization_2015}. In these works, the devil-stick was manipulated by controlling \emph{both} the normal and tangential forces applied on the devil-stick by the actuator. 
They assume that the actuator rolls on the devil-stick without slipping, which causes the contact point to change over time. This further necessitates the assumption that the contact point instantaneously jumps to its original value 
% every time the devil-stick completes a rotation, \emph{i.e.}, for every evolution of the orientation angle of the devil-stick by $2\pi$ rad. 
for every rotation of the devil-stick by $2\pi$ rad.
This resetting of the contact point makes the resulting propeller motion hybrid.

We propose a different approach to realizing stable propeller motion of a devil-stick. In comparison to existing work, the novelty of our approach is as follows:
\begin{itemize}
    \item The control inputs are chosen to be the normal force on the devil-stick, and its location of application.
    \item There is no tangential force applied on the devil-stick by the actuator.
    \item For a stable propeller motion, the control inputs vary smoothly, \emph{i.e.}, there is no instantaneous change in the point of application of the force for every rotation of the devil-stick. 
    % The resulting system is therefore not hybrid.
\end{itemize}

Virtual holonomic constraints (VHCs) \cite{shiriaev_constructive_2005, maggiore_virtual_2013, mohammadi_dynamic_2018} are used to specify a circular trajectory for the center-of-mass of the devil-stick. A continuous feedback linearizing controller enforces the VHCs. A variation of the impulse controlled Poincar\'e map (ICPM) approach \cite{kant_orbital_2020} is used to stabilize the orbit corresponding to a desired propeller motion by 
% varying the applied normal force
applying an additional large-amplitude normal force
when the orientation of the devil-stick crosses a certain value, which makes the overall system hybrid. 
This additional force becomes zero as the desired orbit is stabilized, and the system dynamics corresponding to stable propeller motion is no longer hybrid.
Simulations demonstrate that the presented approach can stabilize a desired propeller motion from arbitrary initial conditions.

\section{System Dynamics} \label{sec2}

\subsection{System Description} \label{sec21}

\begin{figure}[t]
    \centering
    \psfrag{A}{$x$}
    \psfrag{B}{$y$}
    \psfrag{C}{$\ell$}
    \psfrag{D}{$\theta$}
    \psfrag{E}{$m,\, J$}
    \psfrag{F}{\small{center-of-mass $G$}}
    \psfrag{K}{$g$}
    \psfrag{P}{$F$}
    \psfrag{Q}{$r$}
    \includegraphics[width=0.58\linewidth]{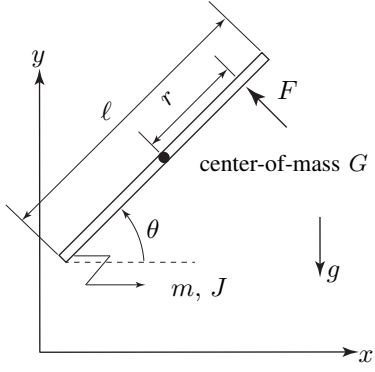}
    \caption{A devil-stick in the vertical plane with configuration variables $(h_x, h_y, \theta)$, and control variables $(F, r)$.}
    \label{fig:description}
\end{figure}

Consider the motion of a 3DOF rigid devil-stick in the vertical $xy$ plane using as shown in Fig. \ref{fig:description}. The configuration of the devil-stick is specified by the three generalized coordinates: $(h_x, h_y, \theta)$, where $h_x$ and $h_y$ denote the Cartesian coordinates of the center-of-mass $G$ of the devil-stick, and $\theta$ denotes the angle it makes with respect to the horizontal $x$ axis, measured positive counterclockwise. 
The devil-stick is manipulated by means of a continuous force $F$ applied normal to the devil-stick at a distance $r$ from $G$. It is assumed that the manipulator applying the force can slide on the devil-stick without friction. Since the number of available control inputs is one fewer than the number of generalized coordinates, the system is underactuated and there is a single passive coordinate. The vector of generalized coordinates is represented as
\begin{equation} \label{eq:generalized-coord}
    q \triangleq \begin{bmatrix} q_1^T & \!\!|\,\,q_2 \end{bmatrix}^T, \quad q_1 = \begin{bmatrix} h_x & h_y \end{bmatrix}^T, \, q_2 = \theta
\end{equation}

\noindent where $q_1 \in \mathbb{R}^2$ are treated as the active coordinates and $q_2 \in \mathbb{R}$ is treated as the passive coordinate.

\begin{remark} \label{rem1}
The coordinate $q_2$ represents a rotation and physically $q_2 + 2\pi \equiv q_2$. However, we have not chosen $q_2 \in S^1$ owing to the structure of the reduced system that results from the choice of VHCs - see 
% the discussion in 
Section \ref{sec24}. 
\end{remark}

\subsection{Dynamic Model} \label{sec22}

The equations of motion of the devil-stick are derived using Lagrange's method. The kinetic and potential energies of the devil-stick are given respectively by:
\begin{equation} \label{eq:kinetic-potential-energy}
    T = \frac{1}{2}m\dot h_x^2 + \frac{1}{2}m\dot h_y^2 + \frac{1}{2}J\dot\theta^2, \qquad
    V = m g h_y 
\end{equation}
% \begin{align}
%     T &= \frac{1}{2}m\dot h_x^2 + \frac{1}{2}m\dot h_y^2 + \frac{1}{2}J\dot\theta^2  \label{eq:kinetic-energy} \\
%     V &= m g h_y \label{eq:potential-energy}
% \end{align}

\noindent from which it follows that the Lagrangian is
\begin{equation} \label{eq:lagrangian}
    L = T - V
      = \frac{1}{2}m\dot h_x^2 + \frac{1}{2}m\dot h_y^2 + \frac{1}{2}J\dot\theta^2 - m g h_y
\end{equation}
% \begin{equation} \label{eq:lagrangian}
% \begin{split}
%     L &= T - V \\
%       &= \frac{1}{2}m\dot h_x^2 + \frac{1}{2}m\dot h_y^2 + \frac{1}{2}J\dot\theta^2 - m g h_y
% \end{split}
% \end{equation}

\noindent To obtain the generalized forces corresponding to each generalized coordinate, we consider the virtual work done by the applied force, given by
\begin{equation} \label{eq:virtual-work}
    \delta W = - F\sin\theta \delta h_x + F\cos\theta \delta h_y + F r \delta\theta
\end{equation}

\noindent The generalized force $Q_j$ corresponding to the generalized coordinate $q_j$, $q_j = h_x, h_y, \theta$, is then given by the term multiplying $\delta q_j$ in \eqref{eq:virtual-work}. Lagrange's equations can be written as
\begin{equation} \label{eq:lagrange}
    \frac{d}{dt} \left( \frac{\partial L}{\partial \dot q_j} \right) - \frac{\partial L}{\partial q_j} = Q_j
\end{equation}

\noindent which, upon simplification, lead to the equations of motion:
\begin{align}
    \ddot h_x &= - \frac{\sin\theta}{m} F \label{eq:eom-hx} \\
    \ddot h_y &= -g + \frac{\cos\theta}{m} F \label{eq:eom-hy} \\
    \ddot\theta &= \frac{1}{J} F r \label{eq:eom-theta} 
\end{align}

\noindent Choosing $u \in \mathbb{R}^2$ as the vector of control inputs given by
\begin{equation} \label{eq:control-inputs}
    u = \begin{bmatrix} F & F r \end{bmatrix}^T
\end{equation}

\noindent the above equations of motion can be rewritten in the standard form \cite{kant_orbital_2020}
\begin{subequations}
\begin{align}
    \ddot{q}_1 =  A(q, \dot q) + B(q) u \\
    \ddot{q}_2 =  C(q, \dot q) + D(q) u
\end{align}
\end{subequations}

\noindent where
\begin{equation}
A = \begin{bmatrix} 0 \\ -g \end{bmatrix},\, 
B = \begin{bmatrix} - \dfrac{\sin q_2}{m} & 0 \\[2ex] \dfrac{\cos q_2}{m} & 0\end{bmatrix},\, 
C = 0,\,
D = \begin{bmatrix} 0 & \dfrac{1}{J} \end{bmatrix}
\end{equation}

\subsection{VHC for Propeller Motion} \label{sec23}

\begin{figure}[t]
    \centering
    \psfrag{A}{$x$}
    \psfrag{B}{$y$}
    \psfrag{D}{$q_2$}
    \psfrag{E}{$\phi$}
    \psfrag{C}{$R$}
    \includegraphics[width=0.58\linewidth]{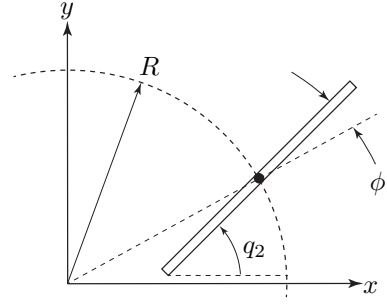}
    \caption{VHC describing the desired trajectory of the center-of-mass of the devil-stick showing the parameters $R$ and $\phi$.}
    \label{fig:vhc}
\end{figure}

We use the following VHC to define the desired trajectory of the center-of-mass of the devil-stick in terms of its orientation:
\begin{equation} \label{eq:VHC-rho}
\rho(q) = q_1 - \Phi(q_2) = 0, \quad \Phi: \mathbb{R} \to \mathbb{R}^2
\end{equation}

\noindent where $\Phi: \mathbb{R} \to \mathbb{R}^2$ is given by
\begin{equation} \label{eq:VHC-Phi}
    \Phi(q_2) = \begin{bmatrix}
        R \cos(q_2 - \phi) \\
        R \sin(q_2 - \phi)
    \end{bmatrix}
\end{equation} 

\noindent This VHC defines the trajectory of the center-of-mass to be along a circle of radius $R$, $R > 0$, with its center at the origin of the $xy$ frame - see Fig. \ref{fig:vhc}. The parameter $\phi$, $\phi \in (-\pi, \pi]$, denotes an offset between the angle subtended by the center-of-mass $G$ at the origin, and the orientation $q_2$. The constraint manifold $\mathcal{C}$ corresponding to the VHC \eqref{eq:VHC-rho} is:
% given by:
\begin{equation} \label{eq:constraint-manifold}
\mathcal{C} = \left\{ (q, \dot q) : q_1 = \Phi(q_2), \dot q_1 = \left[\frac{\partial \Phi}{\partial q_2}\right] \dot q_2 \right\}
\end{equation}

\noindent The continuous control $u_c, u_c \triangleq \begin{bmatrix} F_c & F_c r_c \end{bmatrix}^T$ in \cite{kant_orbital_2020}:
\begin{equation} \label{eq:uc}
\begin{split}
u_c = \left[ B - (\partial \Phi/ \partial q_2) D \right]^{-1} \left[ -A + (\partial^2 \Phi/ \partial q_2^2) \dot q_2^2\right. \\ \left.+ (\partial \Phi/ \partial q_2) C - k_p\rho - k_d\dot\rho \right]
\end{split}
\end{equation}

\noindent where $k_p$ and $k_d$ are positive definite matrices, drives $\rho(t)$ to zero exponentially, enforcing the VHC in \eqref{eq:VHC-rho} and rendering $\mathcal{C}$ controlled invariant.

\begin{remark} \label{rem2}
The control $u_c$ in \eqref{eq:uc} is well-defined and $\mathcal{C}$ is stabilizable $\forall  q_2$ if the matrix 
\begin{equation}
    \left[ B - (\partial \Phi/ \partial q_2) D \right] = \begin{bmatrix}
        -\dfrac{\sin q_2}{m} &  \dfrac{R \sin( q_2 - \phi)}{J} \\[2ex]
         \dfrac{\cos q_2}{m} & -\dfrac{R \cos( q_2 - \phi)}{J}
    \end{bmatrix}
\end{equation}

\noindent is nonsingular \cite[Remark 1]{kant_orbital_2020}, \emph{i.e.}, its determinant
\begin{equation} \label{eq:nonsingular}
    \frac{R}{m J} \sin\phi \neq 0 \quad \Rightarrow \quad \phi \notin\{0, \pi\}
\end{equation}
\vspace{-10pt}
\end{remark}

It follows from \eqref{eq:uc} that the continuous force $F_c$ on the devil-stick, and its point of application $r_c$ are given by
\begin{equation} \label{eq:uc-components}
    F_c = \begin{bmatrix} 1 & 0\end{bmatrix}u_c, \quad r_c =  \begin{bmatrix} 0 & 1\end{bmatrix}u_c/ F_c
\end{equation}

\begin{remark} \label{rem3}
To ensure continuous contact between the actuator and devil-stick, $F_c$ must not change sign. 
On $\mathcal{C}$, \emph{i.e.} when $\rho = 0$ and $\dot\rho = 0$, the value of $F_c$ may be obtained from \eqref{eq:uc} and \eqref{eq:uc-components} as
\begin{equation*}
    F_c = \frac{m R}{\sin\phi} \left[ \dot q_2^2 - \frac{g}{R} \sin(q_2 - \phi) \right]
\end{equation*}
and $F_c$ does not change sign if $\dot q_2^2 > (g/R)\sin(q_2 - \phi)$ $\forall  q_2$. Under this condition, if $\sin\phi$ is positive (negative), the value of $F_c$ is always positive (negative). This requirement also ensures that $r_c$ is always defined since $F_c \neq 0$ for any $q_2$. 
On $\mathcal{C}$, the value of $r_c$ from \eqref{eq:uc} and \eqref{eq:uc-components} is given by
\begin{equation*}
    r_c = \frac{J}{m R} \left[ \frac{\dot q_2^2 \cos\phi - (g/R) \sin q_2}{\dot q_2^2 - (g/R) \sin(q_2 - \phi)} \right]
\end{equation*}
and this expression can be used to ensure that the point of application of the force always lies on the devil-stick, \emph{i.e}, $r_c \in (-\ell/2, \ell/2)$ $\forall q_2$.
\end{remark}

\subsection{Zero Dynamics} \label{sec24}

For system trajectories on the constraint manifold, $\rho(q) \equiv 0$. This implies
\begin{equation}
    q_1 = \Phi(q_2), 
    \dot q_1 = \left[\frac{\partial \Phi}{\partial q_2}\right] \dot q_2, 
    \ddot q_1 = \left[\frac{\partial \Phi}{\partial q_2}\right] \ddot q_2 + \left[\frac{\partial^2 \Phi}{\partial q_2^2}\right] \dot q_2^2
\end{equation}

\noindent from which it follows that 
\begin{subequations} \label{eq:d2h-constraint-manifold}
\begin{align}
    \ddot h_x &= - R\sin( q_2 - \phi) \ddot q_2 - R\cos( q_2 - \phi) \dot q_2^2 \\
    \ddot h_y &=   R\cos( q_2 - \phi) \ddot q_2 - R\sin( q_2 - \phi) \dot q_2^2
\end{align}
\end{subequations}

\noindent Multiplying both sides of \eqref{eq:eom-hx} by $\cos q_2$, both sides of \eqref{eq:eom-hy} by $\sin q_2$, and adding the resulting equations, we can eliminate the input $F$ between them to obtain
\begin{equation} \label{eq:eliminate-F}
    \ddot h_x \cos q_2 + \ddot h_y \sin q_2 = -g \sin q_2
\end{equation}

\noindent Using \eqref{eq:d2h-constraint-manifold} in \eqref{eq:eliminate-F} and simplifying, we obtain the zero dynamics
\begin{equation}
    R\sin\phi\, \ddot q_2 - R\cos\phi\, \dot q_2^2 = - g\sin q_2
\end{equation}

\noindent which can be expressed in the form
\begin{align}
    \ddot q_2 &= \alpha_1( q_2) + \alpha_2( q_2) \dot q_2^2 \label{eq:zero-dynamics-shiriaev} \\
    \alpha_1( q_2) &= -\frac{g \sin q_2}{R \sin\phi}, \quad \alpha_2( q_2) = \cot\phi
\end{align}

\noindent The equation \eqref{eq:zero-dynamics-shiriaev} has an integral of motion \cite{perram_explicit_2003, maggiore_virtual_2013, kant_orbital_2020}
\begin{equation} \label{eq:integral-of-motion}
     E( q_2, \dot q_2) = \frac{1}{2} \mathcal{M}( q_2) \dot q_2^2 + \mathcal{P}( q_2) 
\end{equation}

\noindent where $\mathcal{M}( q_2)$ and $\mathcal{P}( q_2)$ are the mass and potential energy of the reduced system \eqref{eq:zero-dynamics-shiriaev}, given by
\begin{equation} \label{eq:reduced-M}
\begin{split}
     \mathcal{M}( q_2) &= \exp{-2 \int_0^{q_2} \alpha_2(\tau) d\tau} \\
     &= e^{-2  q_2 \cot\phi}
\end{split}
\end{equation}
    
\noindent and
\begin{equation} \label{eq:reduced-P}
\begin{split}
    \mathcal{P}( q_2) &= - \int_0^{q_2} \alpha_1(\tau) \mathcal{M}(\tau) d\tau \\
    &= -\frac{g e^{-2  q_2 \cot\phi} (2 \sin q_2 \cot\phi + \cos q_2)}{R\sin\phi (4\cot^2\phi + 1)}
\end{split}
\end{equation}

\noindent Thus, $E( q_2, \dot q_2)$ can be written as
\begin{equation} \label{eq:integral-of-motion-exp}
    E = e^{-2  q_2 \cot\phi} \left[ \frac{1}{2}\dot q_2^2 - \frac{g (2 \sin q_2 \cot\phi + \cos q_2)}{R\sin\phi (4\cot^2\phi + 1)} \right]
\end{equation}

\noindent It follows from \cite{maggiore_virtual_2013, mohammadi_dynamic_2018} that the zero dynamics \eqref{eq:zero-dynamics-shiriaev} has an Euler-Lagrange structure if and only if $\mathcal{M}$ and $\mathcal{P}$ in \eqref{eq:reduced-M} and \eqref{eq:reduced-P} are $2\pi$-periodic. This condition is met if $\phi \in \{-\pi/2, \pi/2\}$ since for these choices of $\phi$, $\cot\phi = 0 \Rightarrow \exp{-2 q_2 \cot\phi} = 1 \, \forall  q_2$. For these choices of $\phi$, it may be assumed that $q_2 \in S^1$; \eqref{eq:zero-dynamics-shiriaev} then reduces to
\begin{equation}
    \ddot q_2 = 
    \begin{cases}
        &\,\,\,\,\, \dfrac{g \sin q_2}{R}, \quad \phi = -\pi/2 \\[2ex]
        &- \dfrac{g \sin q_2}{R}, \quad \phi =  \pi/2
    \end{cases}
\end{equation}

\noindent which is similar to the dynamics of a simple pendulum. In particular, when $\phi = \pi/2$ the above equation is exactly that of a simple pendulum of length $R$. Further, \eqref{eq:integral-of-motion-exp} reduces to
\begin{equation} \label{eq:integral-of-motion-piby2}
    E = 
    \begin{cases}
        &\dfrac{1}{2}\dot q_2^2 + \dfrac{g \cos q_2}{R}, \quad \phi = -\pi/2 \\[2ex]
        &\dfrac{1}{2}\dot q_2^2 - \dfrac{g \cos q_2}{R}, \quad \phi =  \pi/2
    \end{cases}
\end{equation}

\noindent which indicates stable, periodic behavior, consistent with the results reported in \cite{kawaida_feedback_2003, nakaura_enduring_2004}. The constraint manifold $\mathcal{C}$ is comprised of a dense set of orbits which are stable, but not asymptotically stable \cite{maggiore_virtual_2013, kant_orbital_2020}. The orbits are characterized by the level sets of $E( q_2, \dot q_2)$ in \eqref{eq:integral-of-motion-piby2}. Propeller motion is achieved if the values of $ q_2$ and $\dot q_2$ on $\mathcal{C}$ ensure that the value of $E$ is greater than the maximum value $\mathcal{P}_\mathrm{max}$ of $\mathcal{P}(q_2)$, \emph{i.e.}, $E > g/R$, which corresponds to an orbit for which $\dot q_2$ does not change sign \cite{maggiore_virtual_2013, kant_orbital_2020}.\

For values of $\phi$ satisfying \eqref{eq:nonsingular} chosen such that $\cot\phi \neq 0$, $\mathcal{M}$ and $\mathcal{P}$ are not $2\pi$-periodic, and the system \eqref{eq:zero-dynamics-shiriaev} does not have an Euler-Lagrange structure. Further, it can be seen from \eqref{eq:integral-of-motion-exp} that the system does not permit closed, periodic orbits as the magnitude of $\dot q_2$ changes over every rotation as $q_2$ increments by $2\pi$, because of the term $\exp{-2 q_2\cot\phi}$.

\begin{remark}
    Note that the VHCs in \eqref{eq:VHC-Phi} are not odd since $\Phi(-q_2) \neq -\Phi(q_2)$. If the VHCs were odd, and additional conditions on the mass matrix and potential energy function of the dynamical system were satisfied, the zero dynamics would have an Euler-Lagrange structure for any VHC parameter choices \cite{maggiore_virtual_2013, mohammadi_dynamic_2018, kant_orbital_2020}.
\end{remark}

\begin{remark}
    The zero dynamics \eqref{eq:zero-dynamics-shiriaev} has an Euler-Lagrange structure if $\phi \in \{-\pi/2, \pi/2\}$ even though the VHCs in \eqref{eq:VHC-Phi} are not odd. The corresponding constraint manifold therefore comprises a dense set of orbits which are stable, but not asymptotically stable.
\end{remark}

For the cases when $\phi = \pm \pi/2$, we now describe a procedure to stabilize a desired orbit on the constraint manifold.

\section{Stabilization of Propeller Motion} \label{sec3}

\subsection{Orbit Selection and Poincar\'e Map} \label{sec31}

A distinct propeller motion described by the VHCs \eqref{eq:VHC-Phi}, with $\phi \in \{-\pi/2, \pi/2\}$, is the orbit
\begin{equation} \label{eq:orbit}
    \mathcal{O}^* = \{(q, \dot q) \in \mathcal{C} : E(q_2, \dot q_2) = c^* \}, \quad c^* > g/R
\end{equation}

\noindent We use a modification of the ICPM approach to realize stabilization of $\mathcal{O}^*$. We first define the Poincar\'e section
\begin{equation} \label{eq:poincare-section}
    \Sigma = \{(q, \dot q) \in \mathbb{R}^6 : q_2 \!\!\!\! \mod 2\pi = q_2^*, \dot q_2 > 0\}
\end{equation}

\noindent on which the states are
\begin{equation}
    z = \begin{bmatrix} q_1^T & \dot q^T \end{bmatrix}^T, \quad z \in \mathbb{R}^5
\end{equation}

\noindent If the system trajectory does not lie on $\mathcal{O}^*$, it is assumed that an impulsive force of impulse $I$ is applied when the system trajectory intersects $\Sigma$ without the actuator breaking contact with the devil-stick, and without any discontinuous jumps in $r$ on $\Sigma$. The objective is to design the values of $I$ to drive trajectories to $\mathcal{O}^*$. The dynamics of the impulse-controlled system may be expressed as
\begin{equation} \label{eq:hybrid-map}
    z(k+1) = \mathbb{P} [z(k), I(k)]
\end{equation}

\noindent where $z(k)$ denotes the states on $\Sigma$ immediately prior to application of the impulsive force $I(k)$. 

\subsection{Orbital Stabilization} \label{sec32}

If $(q, \dot q) \in \mathcal{O}^*$, the system trajectory evolves on $\mathcal{O}^*$ under the continuous control $u_c$ in \eqref{eq:uc}. Therefore, $\mathcal{O}^*$ intersects $\Sigma$ at a fixed point $z(k) = z^*$, $I(k) = 0$ of the map $\mathbb{P}$:
\begin{equation} \label{eq:fixed-point}
    z^* = \mathbb{P} (z^*, 0)
\end{equation}

\noindent If $(q, \dot q) \notin \mathcal{O}^*$, $u_c$ ensures convergence of the system trajectory to $\mathcal{C}$, but not necessarily the desired orbit $\mathcal{O}^*$ on $\mathcal{C}$. To stabilize the fixed point $z^*$ on $\Sigma$, and consequently the orbit $\mathcal{O}^*$ \cite{grizzle_asymptotically_2001, nersesov_generalization_2002}, we linearize the map $\mathbb{P}$ about $z(k) = z^*$ and $I(k) = 0$:
\begin{equation} \label{eq:linearized-map}
    e(k+1) = \mathcal{A} e(k) + \mathcal{B} I(k), \quad e(k) \triangleq z(k) - z^* \\
\end{equation} 

\noindent where $\mathcal{A} \in \mathbb{R}^{5 \times 5}$, $\mathcal{B} \in \mathbb{R}^{5}$ are given by:
\begin{align} 
\mathcal{A} &\triangleq  \left[\nabla_{z}\mathbb{P}(z, I)\right]_{z = z^*\!, \,I = 0} \notag \\
\mathcal{B} &\triangleq  \left[\nabla_{I}\mathbb{P}(z, I) \right]_{z = z^*\!, \,I = 0}
\end{align}

\noindent and may be computed numerically \cite{kant_orbital_2020}. The $i$th column $\mathcal{A}_i$ of $\mathcal{A}$ is computed as:
\begin{equation}
    \mathcal{A}_i = \frac{1}{\epsilon_1} \left[ \mathbb{P} (z^* + \delta_i) - z^* \right]
\end{equation}

\noindent where $\delta_i$ is the $i$th column of $\epsilon_1 \mathbb{I}_5$; $\epsilon_1$ is a small number, and $\mathbb{I}_5$ is the $5 \times 5$ identity matrix. To compute $\mathcal{B}$, we observe that a small impulse applied when the system trajectory intersects $\Sigma$ causes the following jump in the states\footnote{Impulsive inputs cause no change in the position coordinates and finite jumps in the velocity coordinates \cite{kant_orbital_2020}.} on $\Sigma$:
\begin{equation}
    S \triangleq \begin{bmatrix} 0 & 0 & \{ B(q_2^*) \eta \}^T & D(q_2^*) \eta \end{bmatrix}^T, \quad \eta \triangleq \begin{bmatrix} \epsilon_2 & \epsilon_2 r^* \end{bmatrix}^T
\end{equation}

\noindent where $r^*$ is the point of application of the force on $\Sigma$ when the system trajectory lies on $\mathcal{O}^*$, obtained from \eqref{eq:uc-components}, and $\epsilon_2$ is a small number. Then, $\mathcal{B}$ is computed as:
\begin{equation}
    \mathcal{B} = \frac{1}{\epsilon_2} \left[ \mathbb{P} (z^* + S) - z^* \right] 
\end{equation}

\noindent If $(\mathcal{A}, \mathcal{B})$ is controllable, the orbit $\mathcal{O}^*$ can be asymptotically stabilized by the discrete feedback
\begin{equation} \label{eq:discrete-feedback}
    I(k) = \mathcal{K} e(k)
\end{equation}

\noindent where $\mathcal{K}$ is chosen to place the eigenvalues of $(\mathcal{A} + \mathcal{B}\mathcal{K})$ inside the unit circle.

The impulsive input $I(k)$ is realized in continuous-time by the high-gain feedback applied in addition to $u_c$
\begin{equation} \label{eq:uhg}
    u_\mathrm{hg} = \begin{bmatrix} F_\mathrm{hg} \\ F_\mathrm{hg} r(k) \end{bmatrix}, \quad 
    F_\mathrm{hg} = \frac{J}{\mu r(k)} [ \dot q_2^\mathrm{des}(k) - \dot q_2 ]
\end{equation}

\noindent where $\mu$ is a small number and
\begin{equation*}
    \dot q_2^\mathrm{des}(k) = \dot q_2(k) + D(q_2^*)
    \begin{bmatrix}
        I(k) \\ I(k) r(k)
    \end{bmatrix} = \dot q_2(k) + \frac{I(k) r(k)}{J}
\end{equation*}

\noindent with $r(k)$ denoting the point of application of the force on $\Sigma$ from \eqref{eq:uc-components}. The high-gain feedback remains active as long as $\abs{ \dot q_2^\mathrm{des}(k) - \dot q_2 } > \epsilon_3$, where $\epsilon_3$ is a small, positive number. At any instant, 
% the force applied on the stick 
$F = F_c + F_\mathrm{hg}$; $F_\mathrm{hg}$ is inactive when the system trajectory is sufficiently close to $\mathcal{O}^*$.

\section{Simulation} \label{sec}

The physical parameters of the devil-stick in SI units are chosen to be:
\begin{equation*}
    m = 0.1, \quad \ell = 0.5, \quad J = \frac{1}{12} m \ell^2 = 0.0021
\end{equation*}

\subsection{Stabilization of a Periodic Orbit}

We consider the VHC given by \eqref{eq:VHC-Phi} with parameters
\begin{equation}
    R = 1, \quad \phi = \pi/2
\end{equation}

\noindent which satisfy \eqref{eq:nonsingular} and define a constraint manifold containing a dense set of stable closed orbits -  see Section \ref{sec24}. The gain matrices $k_p$ and $k_d$ enforcing the VHC were chosen as
\begin{equation} \label{eq:gain-mat}
    k_p = \begin{bmatrix}
        40 & 0 \\ 0 & 40
    \end{bmatrix}, \quad
    k_d = \begin{bmatrix}
        5.5 & 0 \\ 0 & 5.5
    \end{bmatrix}
\end{equation}

\noindent The desired orbit on $\mathcal{C}$ to be stabilized is chosen to be the one passing through
\begin{equation}
    \begin{bmatrix} q^T & \dot q^T \end{bmatrix}^T = \begin{bmatrix} 0 & -1 & 0 & 8 & 0 & 8 \end{bmatrix}^T
\end{equation}

\noindent which is the orbit given by
\begin{equation} \label{eq:orbit-sim}
    \mathcal{O}^* = \{(q, \dot q) \in \mathcal{C} : E(q_2, \dot q_2) = 22.1900 \} 
\end{equation}

\noindent which satisfies the requirement that $E > g/R$ and meets the conditions in Remark \ref{rem3}. 
We choose the following Poincar\'e section on which to implement the controller \eqref{eq:uhg}:
\begin{equation} \label{eq:poincare-section-sim}
    \Sigma = \{(q, \dot q) \in \mathbb{R}^6 : q_2 \!\!\!\! \mod 2\pi = \pi/6, \dot q_2 > 0\}
\end{equation}

\noindent The intersection of $\mathcal{O}^*$ with $\Sigma$ corresponds to the fixed point $z^*$ of the map $\mathbb{P}$ in \eqref{eq:fixed-point}:
\begin{equation}
    \!\! z^* \!=\! \begin{bmatrix}
        0.5000 & -0.8660 & 6.7844 & 3.9170 & 7.8340
    \end{bmatrix}^T
\end{equation}

\begin{figure}[t]
    \centering
    \psfrag{A}{\hspace{-10pt} \footnotesize{$q_2$ (rad)}}
    \psfrag{B}{\hspace{-18pt} \footnotesize{$\dot q_2$ (rad/s)}}
    \psfrag{T}{\hspace{-8pt} \footnotesize{$t$ (s)}}
    \psfrag{Q}{\hspace{-18pt} \footnotesize{$E(q_2, \dot q_2)$}}
    \psfrag{E}{\footnotesize{$E = 22.1900$}}
    \psfrag{F}{\hspace{-10pt} \footnotesize{$F$ (N)}}
    \psfrag{R}{\hspace{-10pt} \footnotesize{$r$ (m)}}
    \psfrag{C}{$\rho_1$}
    \psfrag{D}{$\rho_2$}
    \psfrag{G}{$\dot \rho_1$}
    \psfrag{L}{$\dot \rho_2$}
    \psfrag{O}{$\mathcal{O}^*$}
    \psfrag{H}{\footnotesize{$E = 18.4408$}}
    \includegraphics[width=\linewidth]{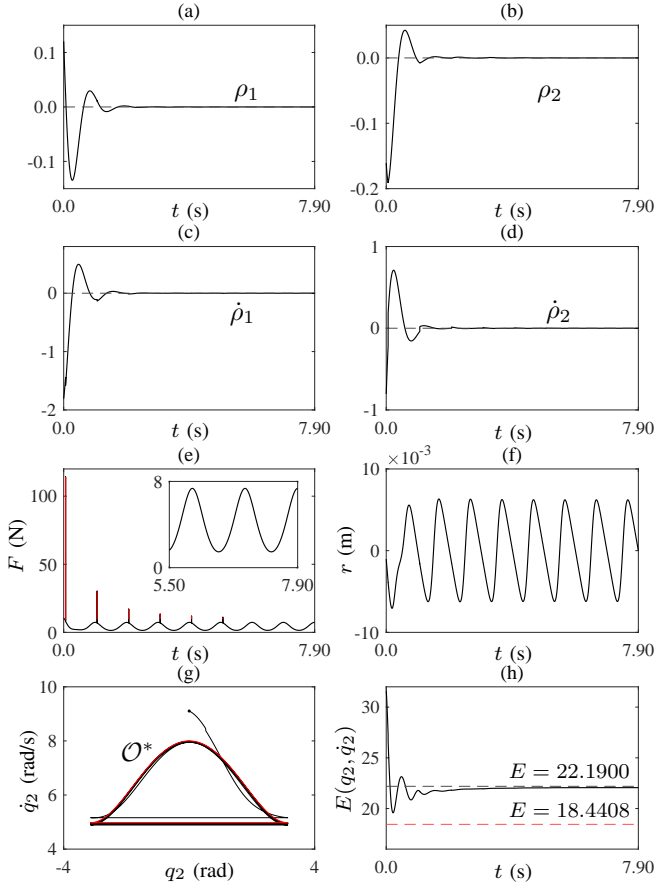}
    \caption{Stabilization of propeller motion of a devil-stick from arbitrary initial conditions: 
    (a)-(b) show the components of $\rho(q)$, (c)-(d) show the components of $\dot \rho(q)$,
    (e) shows the applied normal force $F$, with the high-gain forcing shown in red; the inset shows a magnified view of the continuous force between $t = 5.50$ s and $t = 7.90$ s, (f) shows the point of application $r$ of the force,
    (g) shows the phase portrait of the passive coordinate $q_2$, and (h) shows the value of $E(q_2, \dot q_2)$ from \eqref{eq:integral-of-motion-piby2}.}
    \label{fig:sim-orbit}
\end{figure}

\noindent The values of $\mathcal{A}$ and $\mathcal{B}$ were obtained numerically as
\begin{equation*}
\begin{split}
    \mathcal{A} &= \begin{bmatrix}
    0.0309 & 0.0000 & -0.0075 &  0.0000 & 0.0065 \\
    0.0000 & 0.0310 &  0.0000 & -0.0075 & 0.0038 \\
    4.4249 & 2.3867 &  0.2962 &  1.0266 & 0.0963 \\
    2.3802 & 1.6797 &  0.1292 &  0.6652 & 0.0556 \\
    4.7604 & 2.7559 &  0.2583 &  1.1854 & 0.1837
    \end{bmatrix}
    \\ 
    \mathcal{B} &= \begin{bmatrix}
    0.0338 & -0.0694 & 7.3592 & 5.0882 & 8.8663
    \end{bmatrix}^T
\end{split}
\end{equation*}

\noindent The eigenvalues of $\mathcal{A}$ do not all lie within the unit circle. The pair $(\mathcal{A}, \mathcal{B})$ is controllable, and the gain matrix $\mathcal{K}$ in \eqref{eq:discrete-feedback} was obtained as:
\begin{equation*}
    \mathcal{K} = \begin{bmatrix}
    -0.5406 & -0.3149 & -0.0318 & -0.1335 & -0.0163
    \end{bmatrix}
\end{equation*}

\noindent using LQR with the gain matrices % $\mathcal{Q} = \mathbb{I}_5$ and $\mathcal{R} = 2$.
\begin{equation*}
    \mathcal{Q} = \mathbb{I}_5, \quad \mathcal{R} = 2
\end{equation*}

The initial condition for simulation is chosen to be
\begin{equation} \label{eq:initial-periodic}
\begin{split}
    &\begin{bmatrix} q^T(0) & \dot q^T(0) \end{bmatrix}^T \\= 
    &\begin{bmatrix} 0.1206 & -1.1608  &       0 & 7.2965 & -0.8040 & 9.1055 \end{bmatrix}^T
\end{split}
\end{equation}

\begin{figure}[t]
    \centering
    \psfrag{X}{\hspace{-10pt} \footnotesize{$h_x$ (m)}}
    \psfrag{Y}{\hspace{-10pt} \footnotesize{$h_y$ (m)}}
    \includegraphics[width=0.72\linewidth]{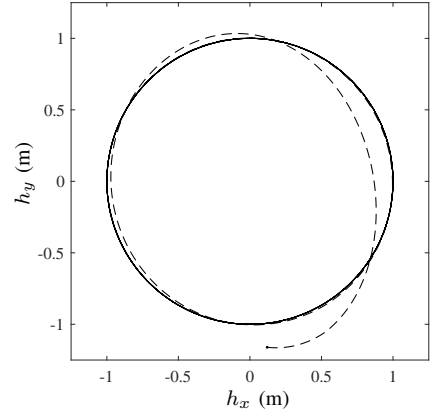}
    \caption{Trajectory of the center-of-mass of the devil-stick.}
    \label{fig:trajectory}
\end{figure}

\noindent which does not lie on $\mathcal{O}^*$, and is not on the constraint manifold $\mathcal{C}$. The simulation results are shown in Fig. \ref{fig:sim-orbit} for $8$ rotations of the devil-stick, corresponding to a duration of approx. $7.90$ s. The high-gain feedback \eqref{eq:uhg} is realized with $\mu = 0.0005$ and $\epsilon_3 = 0.001$. The components of $\rho(q)$ and $\dot \rho(q)$ are shown in Fig. \ref{fig:sim-orbit}(a)-(b) and Fig. \ref{fig:sim-orbit}(c)-(d) respectively, demonstrating convergence of the system trajectory to $\mathcal{C}$. The normal force $F$ and its point of application $r$ are shown in Fig. \ref{fig:sim-orbit}(e) and Fig. \ref{fig:sim-orbit}(f) respectively. It can be seen that no high-gain inputs are applied for $k > 6$ as the system trajectory is sufficiently close to $\mathcal{O}^*$. The phase portrait of the passive coordinate $q_2$ is shown in Fig. \ref{fig:sim-orbit}(g), with $q_2$ plotted in the range $(-\pi, \pi]$. The desired orbit $\mathcal{O}^*$ is shown in red, and it can be seen that the system trajectory asymptotically converges to $\mathcal{O}^*$. 
The value of $E(q_2, \dot q_2)$ from \eqref{eq:integral-of-motion-piby2} is shown in Fig. \ref{fig:sim-orbit}(h). As the desired orbit $\mathcal{O}^*$ is stabilized, $E$ settles at its value in \eqref{eq:orbit-sim}. In the absence of the high-gain controller, \emph{i.e.}, under the action of $u_c$ alone, the system trajectory settles to the orbit with $E = 18.4408$ from the same initial conditions; this value is shown in red in Fig. \ref{fig:sim-orbit}(h). The trajectory of the center-of-mass of the devil-stick is shown in Fig. \ref{fig:trajectory}.

\begin{remark}
    For the choice of orbit in \eqref{eq:orbit-sim}, gain matrices in \eqref{eq:gain-mat} and initial conditions \eqref{eq:initial-periodic}, it is observed that $F > 0$ % and $r_c \in (-\ell/2, \ell/2)$ 
    $\forall t$ since the condition in Remark \ref{rem3} is met, and the value of $I(k) > 0 \, \forall \, k = 1, 2, \dots, 6$. The approach presented here does not explicitly constrain the control inputs, and the requirement that $F$ does not change sign may be violated if the initial conditions require $I(k)$ to be negative. This limitation may be addressed by using a control strategy such as MPC which imposes input constraints.
\end{remark}

\subsection{Aperiodic Motion}

\begin{figure}
    \centering
    \psfrag{A}{\hspace{-10pt} \footnotesize{$q_2$ (rad)}}
    \psfrag{B}{\hspace{-18pt} \footnotesize{$\dot q_2$ (rad/s)}}
    \psfrag{T}{\hspace{-8pt} \footnotesize{$t$ (s)}}
    \psfrag{Q}{\hspace{-18pt} \footnotesize{$E(q_2, \dot q_2)$}}
    \psfrag{E}{\hspace{-5pt} \footnotesize{$E = 22.1934$}}
    \psfrag{F}{\hspace{-10pt} \footnotesize{$F$ (N)}}
    \psfrag{R}{\hspace{-10pt} \footnotesize{$r$ (m)}}
    \includegraphics[width=\linewidth]{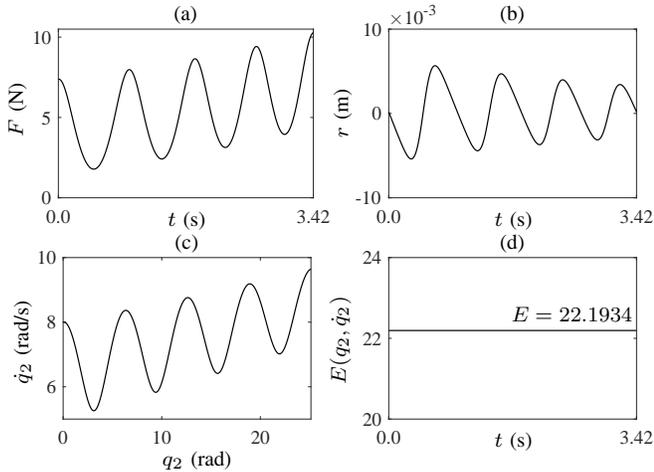}
    \caption{Aperiodic motion of a devil-stick:
    (a) shows the applied normal force $F$, (b) shows the point of application $r$ of the force,
    (c) shows the phase portrait of the passive coordinate $q_2$, and (d) shows the value of $E(q_2, \dot q_2)$ from \eqref{eq:integral-of-motion-exp}.}
    \label{fig:sim-aperiodic}
\end{figure}

We present a simulation with the VHC \eqref{eq:VHC-Phi} and parameters
\begin{equation}
    R = 1, \quad \phi = \pi/2 - 0.01
\end{equation}

\noindent which also satisfy \eqref{eq:nonsingular}. However, for this value of $\phi$, the reduced dynamics \eqref{eq:zero-dynamics-shiriaev} does not have an Euler-Lagrange structure, and the constrant manifold does not possess closed, periodic orbits - see Section \ref{sec24}. 
% The same gain matrices as \eqref{eq:gain-mat} were used to enforce the VHC. 

We simulate the behavior of the system on the constraint manifold $\mathcal{C}$ under the action of $u_c$ alone from the initial conditions
\begin{equation}  \label{eq:initial-aperiodic}
    \begin{bmatrix} q^T(0) & \dot q^T(0) \end{bmatrix}^T = \begin{bmatrix} 0 & -1 & 0 & 8 & 0 & 8 \end{bmatrix}^T
\end{equation}

\noindent The value of $E(q_2, \dot q_2) = 22.1934$ for these initial conditions. This value of $E$ is slightly different from the value of $E$ in \eqref{eq:orbit-sim} since the value of $\phi$ is slightly different from $\pi/2$. The simulation results are shown in Fig. \ref{fig:sim-aperiodic} for $4$ rotations of the devil-stick, corresponding to a duration of approx. $3.42$ s. Since the system trajectory evolves on $\mathcal{C}$, $\rho(q) \equiv 0$, and the components of $\rho(q)$ and $\dot \rho(q)$ are not plotted. The values of $F$ and $r$ are plotted in Fig. \ref{fig:sim-aperiodic}(a) and Fig. \ref{fig:sim-aperiodic}(b), showing their aperiodic nature. The phase portrait of the passive coordinate $q_2$ is shown in Fig. \ref{fig:sim-aperiodic}(c), showing the increase in $\dot q_2$ over every rotation. It can be seen from Fig. \ref{fig:sim-aperiodic}(d) that the value of $E(q_2, \dot q_2)$ from \eqref{eq:integral-of-motion-exp} remains steady at its original value.

\section{Conclusion} \label{sec:conclusion}

A novel method for realizing rotary propeller motion of a devil-stick is presented. In contrast to previous works, which control the normal and tangential force applied to the devil-stick, our approach controls the normal force and its point of application on the devil-stick by removing friction between the actuator and devil-stick.
The resulting propeller motion in our approach is then such that the contact point between the actuator and devil-stick is not reset after every rotation of the devil-stick. Virtual holonomic constraints are used to describe a family of stable propeller motions, each with different speeds of rotation. A variation of the impulse controlled Poincar\'e map approach, wherein intermittent impulsive forces are applied to the devil-stick without loss of contact, is employed to stabilize a propeller motion with a desired speed of rotation. As the system trajectory converges to the desired propeller motion, the magnitude of the impulsive forces converge to zero, and stability of the trajectory is guaranteed under continuous control alone. Simulation results demonstrate the efficacy of the proposed approach in stabilizing propeller motion from arbitrary initial conditions. Future work will focus on control design with explicit input constraints, and experimental validation.

%\appendices

% \balance
\bibliographystyle{IEEEtran}      % basic style, author-year citations
\bibliography{ref}   % name your BibTeX data base

\end{document}